\documentclass[letterpaper]{article}

\usepackage{natbib,alifeconf}  

\title{How Mutation Alters Fitness of Cooperation in Networked Evolutionary Games}
\author{Genki Ichinose$^{1}$, Yoshiki Satotani$^{2}$ \and Hiroki Sayama$^3$ \\
\mbox{}\\
$^1$Shizuoka University, Hamamatsu, 4328561, JAPAN \\
$^2$Okayama University, Okayama, 7008530, JAPAN \\
$^3$Binghamton University, Binghamton, NY 13902 , USA\\
ichinose.genki@shizuoka.ac.jp} 

\begin{document}
\maketitle

\begin{abstract}
Cooperation is ubiquitous in every level of living organisms.
It is known that spatial (network) structure is a viable mechanism for cooperation to evolve.
Until recently, it has been difficult to predict whether cooperation can evolve at a network (population) level.
To address this problem, Pinheiro et al.~proposed a numerical metric, called Average Gradient of Selection (AGoS) in 2012. AGoS can characterize and forecast the evolutionary fate of cooperation at a population level.
However, stochastic mutation of strategies was not considered in the analysis of AGoS.
Here we analyzed the evolution of cooperation using AGoS where mutation may occur to strategies of individuals in networks.
Our analyses revealed that mutation always has a negative effect on the evolution of cooperation regardless of the fraction of cooperators and network structures.
Moreover, we found that mutation affects the fitness of cooperation differently on different social network structures.
\end{abstract}

\section{Introduction}
Cooperation is ubiquitous in every level of living organisms and has played an important role in the major evolutionary transitions \citep{West18082015, Michod15052007, Michod13062006}.
In principle, cooperators benefit others by incurring some costs to themselves, while defectors do not pay any costs.
Therefore, cooperation cannot be an evolutionarily stable strategy for a noniterative game in a well-mixed population \citep{nowak06evolutionaryDynamicsBOOK, MaynardSmith1973, TaylorJonker1978, MaynardSmithBOOK1982, HofbauerBOOK1998}.

In such a situation, spatial (network) structure is a viable mechanism for cooperation to evolve \citep{NowakMay1992, SantosPacheco1995}.
However, until recently, it has been difficult to predict whether cooperation can evolve at a network (population) level due to the complex interactions between evolution of strategies and topologies of networks.
To address this problem, \citet{PinheiroPachecoSantos2012} proposed a numerical metric, called Average Gradient of Selection (AGoS), to characterize and forecast the evolutionary fate of cooperation at a population level.
AGoS can analyze the dynamics of the evolution of cooperation in structured populations even when nontrivial selection pressure is introduced \citep{PinheiroSantosPacheco2012}, when nonlinear imitation probability is used \citep{Dai_etal2013}, and also when structures and states of networks change over time (adaptive social networks) \citep{PinheiroSantosPacheco2016}.

In these earlier studies, however, stochastic mutation of strategies was not considered.
It is important to incorporate such mutation because they frequently occur in real societies and also because results obtained with stochastic fluctuations of strategies would provide more robust observations and conclusions.
Here we analyze the evolution of cooperation using AGoS where mutation may occur to strategies of individuals in networks.

\section{Model}
We developed an agent-based model for the analysis of AGoS.
Individuals are placed on the nodes in a network and they interact only with their neighbors.
The networks used in this paper are described in the next subsection.
Each individual can take one of two strategies: Cooperation ($C$) or Defection ($D$).

\begin{figure}[t]
\begin{center}
\includegraphics[width=7cm]{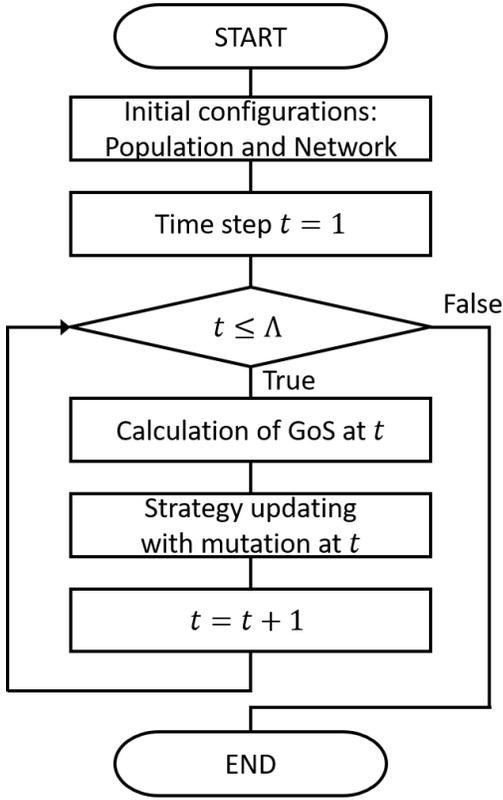}
\caption{Flowchart of the model for each simulation $r$. Simulations are repeated $\Omega$ times in total. AGoS is calculated by using those whole information.}
\label{model}
\end{center}
\end{figure}

Once the composition of individuals in a network is given, we can calculate the probability of increasing or decreasing the number of cooperators by one, called Gradient of Selection (GoS) at time $t$ \citep{PinheiroPachecoSantos2012}.
Simultaneously, we can also update the strategies of individuals based on the framework of evolutionary games at time $t$.
Right after the strategy updating, mutation (flipping strategy from $C$ to $D$ or vice versa) occurs with probability $m$.
Therefore, in evolutionary simulations, the calculations of GoS and the strategy updating with mutation take place alternately in one time step and these processes are repeated.
The calculation of GoS and the strategy updating with mutation are repeated $\Lambda$ time-steps.
Figure \ref{model} gives the flow of the model for each simulation.
Simulations are repeated $\Omega$ times in total.

\subsection{Network structure}
\citet{PinheiroPachecoSantos2012} revealed that cooperation was sustained in a network level and network structures led to the different evolutionary results of cooperation.
Following to the existing work \citep{PinheiroPachecoSantos2012, PinheiroSantosPacheco2012}, we focus on two classes of network structures: homogeneous and heterogeneous.
We use homogeneous in the sense that every individual has the same degree. In the case of heterogeneous, individuals can have different degrees.

For homogeneous networks, we use two types: homogeneous random networks (HR) \citep{SantosRodriguesPacheco2005} and square lattice (Lattice). HR is created by randomizing links from homogeneous regular ring networks.
For heterogeneous networks, we use Barab\'asi-Albert scalfe-free networks (BA) \citep{BarabasiAlbert1999}. In BA, a small number of nodes called hubs connect with a substantial number of links while most other nodes connect with a few nodes.

In the initial setting, the number of cooperators in a population, $j \in [0, N]$, is given to each simulation. The locations of $j$ cooperators and $N-j$ defectors in a network are random. 

\subsection{AGoS calculation}
Once each simulation starts, GoS of the given population at time $t$ is calculated.
If there is no population (network) structure, the probability of  the change of cooperators can be calculated analytically.
However, in structured populations, the calculation is difficult due to the complex connections on networks.

\begin{figure}[t]
\begin{center}
\includegraphics[width=7cm]{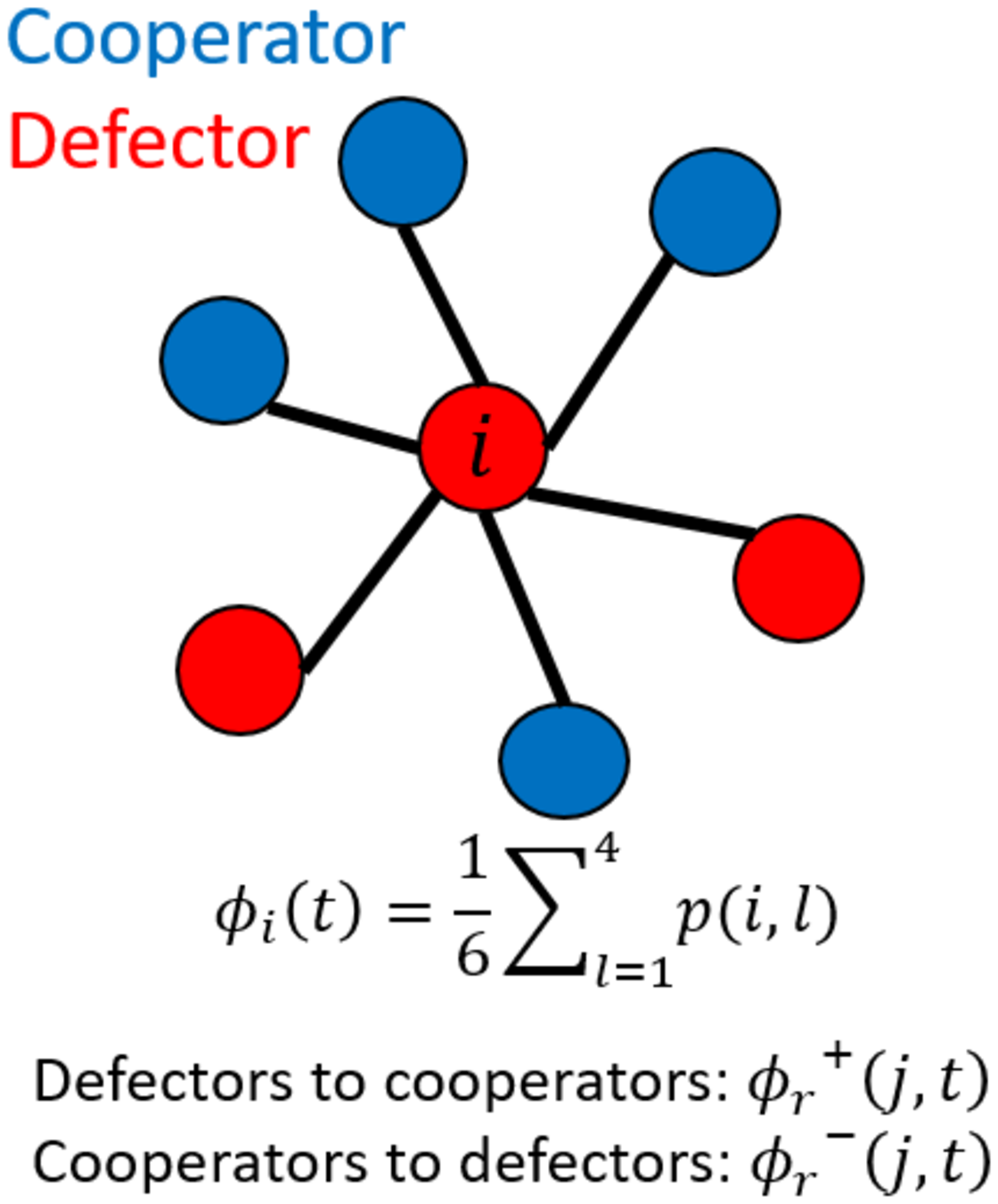}
\caption{Example of $\phi_i (t)$. In this case, the center individual $i$ has $k_i =6$ neighbors and $\bar{n}_i =4$ different strategies in the neighbors. The probability that $i$ ($D$) changes to the other strategy ($C$) is $\phi_i (t)=\frac{1}{6} \sum_{l=1}^{4} p(i, l)$. $\phi_i (t)$ can be classified into two variables: $\phi_{r}^{+} (j,t)$ or $\phi_{r}^{-} (j,t)$. $\phi_{r}^{+} (j,t)$ represents the change from defectors to cooperators. The opposite case is represented by $\phi_{r}^{-} (j,t)$.}
\label{phi}
\end{center}
\end{figure}

The GoS gives the numerical solution for the evolution of cooperation even in such a situation.
Let $\phi_i (t)$ be the probability that $i$'s strategy changes to the other different strategy (from $C$ to $D$ or from $D$ to $C$).
This can be the product of two terms: the probability of selecting a neighbor with the different strategy, $\frac{\bar{n}_i}{k_i}$, and the average probability that $i$ imitates the different strategy, $\frac{1}{\bar{n}_i} \sum_{l=1}^{\bar{n}_i} p(i, l)$, where $k_i$ is the number of $i$'s neighbors and $\bar{n}_i$ is the number of neighbors which has the different strategy opposite to $i$'s. $p(i, l)$ is defined in the next subsection.
Thus, $\phi_i (t)$ can be defined as follows. An example of $\phi_i (t)$ is illustrated in Fig.~\ref{phi}.

\begin{equation}
\phi_i (t)=\frac{\bar{n}_i}{k_i}\frac{1}{\bar{n}_i}\sum_{l=1}^{\bar{n}_i} p (i, l)=\frac{1}{k_i}\sum_{l=1}^{\bar{n}_i} p (i, l).
\end{equation}

From the definition, for a given simulation $r$, the probability to increase the number of $C$ at $j$ (the number of $C$s in the network) is $\phi_{r}^{+} (j,t)=\frac{1}{N}\sum_{i \in s_D}\phi_i (t)$ while the probability for the decrease of the number of $C$s is $\phi_{r}^{-}(j, t) =\frac{1}{N}\sum_{i \in s_C} \phi_i (t)$, where $N$ is the population size, $s_D$ is a set of defectors ($|s_D|=N-j$), and $s_C$ is a set of cooperators ($|s_C|=j$).
Then, GoS at time $t$ in a given simulation $r$ is defined as the difference between them, as

\begin{equation}
G_r (j, t)=\phi_{r}^{+}(j, t) - \phi_{r}^{-}(j, t).
\end{equation}

Finally, we obtain the Average Gradient of Selection (AGoS) \citep{PinheiroPachecoSantos2012}, which averages the GoS by the number of time steps $\Lambda$ and the number of simulations $\Omega$, as

\begin{equation}
G^{A} (j)=\frac{1}{\Lambda \Omega}\sum_{r=1}^{\Omega}\sum_{t=1}^{\Lambda} G_r (j,t).
\end{equation}

\subsection{Strategy updating with mutation}
After the GoS calculation, the strategy updating takes place as follows.
In each time step, we randomly choose one individual $i$ from the population.
The individual plays the Prisoner's Dilemma (PD) game with its $k_i$ neighbors and accumulates the payoffs resulting from the games.
In each game, both individuals obtain payoff $R$ for mutual cooperation while $P$ for mutual defection.
If one selects cooperation while the other does defection, the former obtains the sucker's payoff $S$ while the latter obtains the highest payoff $T$, the temptation to defect.
The relationship of the four payoffs is usually $T > R > P > S$ in PD games.
AGoS is used for the other types of collective games including Stag Hunt \citep{Pacheco_etal2009} and Snowdrift \citep{Santos_etal2012} games.

Following the parameter settings used in the model by \citet{PinheiroPachecoSantos2012, PinheiroSantosPacheco2012}, we used $P = 0, R = 1$, and $S = 1-T$, while $T > 1$.
The neighbors of $i$ also play the PD game with their neighbors and accumulate the payoffs.
Let $\pi_i$ and $\pi_l$ be the payoffs of individual $i$ and $l$ (one of the randomly selected $i$'s neighbors), respectively.
Based on the framework of evolutionary games, higher fitness will be imitated more.
In order to realize this, we use the pairwise comparison rule \citep{TraulsenNowakPacheco2006, TraulsenPachecoNowak2007, SzaboToke1998}.
Individual $i$ imitates $l$'s strategy with the probability 

\begin{equation}
p(i, l) = [1+\mathrm{e}^{-\beta (\pi_l -\pi_i)}]^{-1},
\end{equation}
where $\beta \geq 0$ controls the intensity of selection. For $\beta=0$, there is no selection pressure, meaning that evolutionary dynamics proceeds by random drift. As $\beta$ becomes larger, the tendency that strategies with higher payoffs will be imitated increases.

In the previous studies \citep{PinheiroPachecoSantos2012, PinheiroSantosPacheco2012}, the next time step immediately follows after the strategy updating.
In contrast, we incorporate mutation in this paper. Specifically, the strategy of individual $i$ changes to the different strategy with probability $m$ after the strategy updating.
We focus on how mutation alters fitness of cooperation in networks.

\section{Results}
In each simulation, we calculated the GoS and conducted the strategy updating with mutation every time step, which was iterated $\Lambda=10^5$ time steps.
The population was composed of $N=1000$ ($N=961$ for Lattice) individuals and an average degree was $k=4$. The intensity of selection was $\beta = 10.0$.
We ran $\Omega=30 \times (N+1)$ simulations in total (we ran 30 runs for each $j$ from $j=0$ to $j=N$).
Thus, we set $\Omega=30030$ for HR and BA while $\Omega=28860$ for Lattice.
The temptation to defect $T$ was different depending on networks.
We varied the values of  mutation probability $m$ as the experimental parameter.  

\subsection{Case of HR networks}
\begin{figure}[t]
\includegraphics[width=\columnwidth]{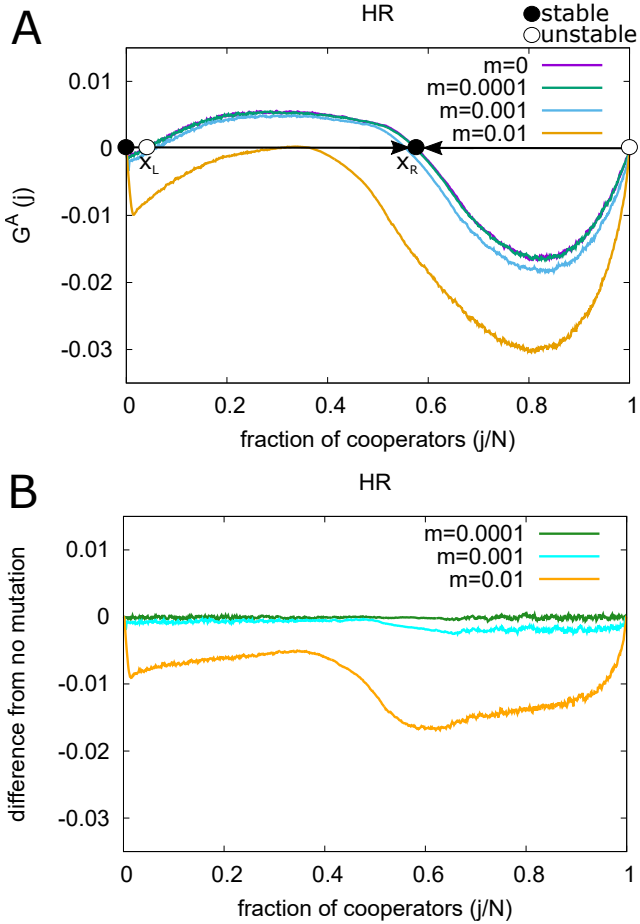}
\caption{(A) AGoS on homogeneous random networks (HR). $T=1.005$. HR has two unstable equilibrium points ($x_L$ and 1) and two stable equilibrium points ($x_R$ and 0). (B) Difference of AGoS from no mutation in HR.}
\label{AGoS_HR}
\end{figure}

We first see the results of homogeneous networks.
Figure \ref{AGoS_HR}A shows the AGoS ($G^A (j)$) on HR networks where the mutation probabilities $m$ are varied.
When $m=0$, the result perfectly matches the corresponding case of \citet{PinheiroPachecoSantos2012}.

It has two stable equilibrium points and two unstable equilibrium points.
One of the unstable points, $x_L$, exists at $j/N =0.039$. One of the stable points, $x_R$, exists at $j/N=0.576$.
The other unstable and stable points exist at $j/N =0$ and $j/N =1$, respectively.
As the arrows suggest, as long as $j/N > x_L$, the population composition converges to the stable point $x_R$.
Thus, unlike the case of well-mixed populations, cooperation and defection can co-exist in the network.
In other words, from a global, population-level perspective, HR networks can sustain cooperation even though all individuals play PD.
As $m$ becomes larger, $G^A (j)$ goes down overall. Thus, the stable coexistence point $x_R$ becomes lower.
In the case of $m>10^{-2}$, we expect that the only stable point exists at $j/N=0$ because $G^A (j)$ is always likely to be lower than 0, meaning that cooperation can no longer exist.
Therefore, mutation is always harmful for cooperation although mutation is neutral itself.

Figure \ref{AGoS_HR}B shows the difference of $G^A (j)$ with mutations on HR networks when we set the case of no mutation $m=0$ as the baseline.
Compared to the lower $j$, the higher $j$ leads to the lower $G^A (j)$ more as the mutation probability increases.
This is because, in the case of the higher $j$, clusters of cooperators are easily destroyed by the mutation.

\subsection{Case of lattice networks}
\begin{figure}[t]
\includegraphics[width=\columnwidth]{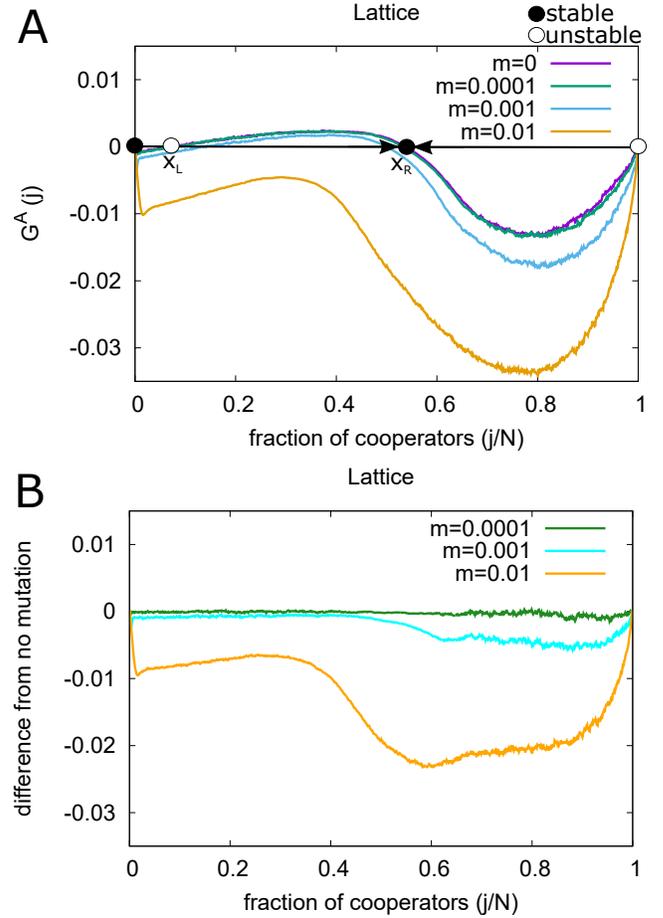}
\caption{(A) AGoS on square lattice (Lattice). $T=1.005$. Similar to HR, Lattice has two unstable equilibrium points ($x_L$ and 1) and two stable equilibrium points ($x_R$ and 0). (B) Difference of AGoS from no mutation in Lattice.}
\label{AGoS_Lattice}
\end{figure}

Figure \ref{AGoS_Lattice}A shows  the AGoS on lattice networks where the mutation probabilities $m$ are varied.
The tendency of this result is the same with the HR networks because lattice is classified into homogeneous networks.

It has also two stable equilibrium points and two unstable equilibrium points.
One of the unstable points, $x_L$, exists at $j/N =0.068$. One of the stable points, $x_R$, exists at $j/N=0.537$.
Hence, the density of co-existence is lower than HR networks ($0.537<0.576$).
The other unstable and stable points exist at $j/N =0$ and $j/N =1$, respectively.
If $j/N > x_L$, cooperation and defection can co-exist in the network.
As $m$ becomes larger, $G^A (j)$ goes down overall. Thus, the stable coexistence point $x_R$ becomes lower.
When $m=10^{-2}$, the only stable point exists at $j=0$. In this case, cooperation finally converges to 0 (extinction) no matter the initial fraction of cooperators ($j/N$) because $G^A (j)$ is always lower than 0.

Figure \ref{AGoS_Lattice}B shows the difference of $G^A (j)$ with mutations on lattice networks when we set the case of no mutation $m=0$ as the baseline.
The value of AGoS is small in the positive zone ($G^A (j)>0$) while the absolute value $|G^A|$ in the negative zone ($G^A (j)<0$) is large when $m=0.01$ (See Fig.~\ref{AGoS_Lattice}A).
In lattice networks, both cooperation and defection are localized. Therefore, when the number of cooperators is low,  only small clusters of cooperators can exist. It is difficult for them to expand their regions because they do not affect faraway places.
In contrast, when the number of cooperators is high, there can be some big cooperative clusters. In this case, if mutation to defection takes place in the clusters, it immediately changes cooperative clusters to defectors. Thus, the absolute value of AGoS  $|G^A|$ when $m=0.01$ in the negative zone may become large.

\subsection{Case of BA networks}
\begin{figure}[t]
\includegraphics[width=\columnwidth]{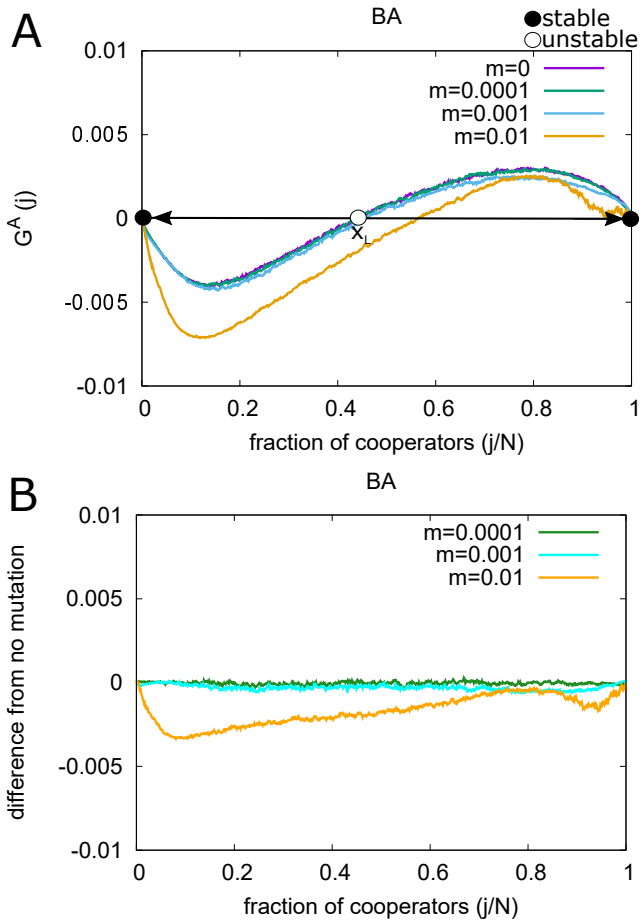}
\caption{(A) AGoS on BA scale-free networks (BA). $T=1.25$. Unlike homogeneous networks (HR and Lattice), BA has one unstable equilibrium points ($x_L$) and two stable equilibrium points (0 and 1). (B) Difference of AGoS from no mutation in BA.}
\label{AGoS_BA}
\end{figure}

Figure \ref{AGoS_BA}A shows  the AGoS on BA networks where the mutation probabilities $m$ are varied.
When $m=0$, there are one unstable equilibrium point $x_L=0.441$ and two stable equilibrium points, $j/N=0$ and $j/N=1$.
This is the same with the one observed in the previous study \citep{PinheiroPachecoSantos2012}.
This means that cooperation becomes dominant when $j/N > x_L$.
As $m$ increases, the number of cooperators needed for sustaining the dominance of cooperation becomes larger.
However, it is still possible for cooperation to become dominant even if the mutation rate is high ($m=0.01$) in BA networks.

Figure \ref{AGoS_BA}B shows the difference of $G^A (j)$ with mutations on BA networks when we set the case of no mutation $m=0$ as the baseline.
Interestingly, mutation is especially harmful for cooperation in the lower $j$ in BA networks, which is the opposite result compared to homogeneous networks (HR and lattice).
In the case of BA networks, cooperative hubs surrounded by cooperators are robust to defectors' invasion.
Thus, cooperation is not so affected when $j$ is large.

\section{Conclusion}
In this paper, we analyzed the evolution of cooperation at a population level in the presence of mutation by the AGoS.
We used two classes of networks: homogeneous (HR and lattice) and heterogeneous (BA).
Our analyses revealed that mutation always has a negative effect on cooperation regardless of the fraction of cooperators and network structures, because local clusters of cooperators can easily be destroyed by mutation.

Interestingly, we found that mutation is particularly harmful to cooperation when the fraction of cooperation is high in homogeneous networks (HR and lattice), but so it is when the fraction of cooperation is low in heterogeneous networks (BA).
This may be due to that hubs surrounded by cooperators are robust to mutated defectors.
If we assume average payoffs rather than accumulated payoffs as considered here, we may have different results, as previously suggested \citep{IchinoseSayama2017}.
We also may have different results when the game structure is different (e.g. the Snowdrift game) even when the network structures are the same, as \citet{HauertDoebeli2004} have revealed.

These results indicate the importance of considering random noise (mutation), which was largely overlooked in the literature, in studying the evolution of cooperative behavior in social networks.
Mutation can be considered genetic changes if we assume biological systems. If we assume cultural systems, mutation can be considered a stochastic behavior to explore new behaviors.
Although we showed such a random exploration was harmful to keep cooperative societies, if we consider different forms of exploration, those explorations may work beneficially for societies, such as collaborative problem solving \citep{SayamaDionne2015}.
We would like to consider those cases in the future work.

\section{Acknowledgements}

G.I. acknowledges the support by Hayao Nakayama Foundation For Science \& Technology \& Culture.

\footnotesize

\end{document}